\documentstyle[aps]{revtex}
\begin{document}
\begin{center}
ORIENTATIONAL MELTING OF TWO-SHELL CARBON NANOPARTICLES:
MOLECULAR DYNAMICS STUDY.\\

{\it Yu. E. Lozovik\footnote{Corresponding author.
Fax: +7-095-334-0886; e-mail: lozovik@isan.troitsk.ru}, A. M. Popov}

{\it Institute of Spectroscopy, Russian Academy of Science, 142190,\\
Troitsk, Moscow region, Russia}
\end{center}

\begin{abstract}
The energetic characteristics of two-shell carbon nanoparticles ("onions")
with different shapes of second shell are calculated.
The barriers of relative rotation of shells
are found to be surprisingly small; therefore,
free relative rotation of shells can take place at room temperature.
The intershell orientational melting of the nanoparticle $C_{60}@C_{240}$
is studied by molecular dynamics. The parameters of Arrhenius formula for
jump rotational intershell diffusion are calculated.
The definition of orientational melting temperature is proposed
as the temperature when the transition probability over barrier
between equivalent potential minima is equal to 1/2.
The temperature of orientational melting of the nanoparticle
$C_{60}@C_{240}$ is about 60 K.
\end{abstract}

\section{Introduction}

The discovery of fullerenes \cite{0} and the elaboration on method
of their production in arc discharge \cite{5} give rise to
interest in another carbon nanostructures produced in arc
discharge, in particular, nanoparticles with shell structure
\cite{6,12}. A set of works is devoted to studying their structure and
energetics \cite{pe}--\cite{p6}. Nevertheless, attention has
not yet been given to thermodynamical properties of carbon
nanoparticles with shell structure.

The melting of a single cluster can differ essentially from phase
transitions in macroscopic systems \cite{1}-\cite{4a}. Particularly,
the melting of a mesoscopic cluster with shell structure can manifest
itself as a hierarchy of rearrangements with breaking intershell
orientational order and then breaking shell structure and order in
particles positions inside shells. For example, in 2D mesoscopic clusters
with Coulomb \cite{1}-\cite{shv}, screened Coulomb \cite{1a},
logarithmic \cite{4} and dipole \cite{4a,4b} interaction between
particles, the orientational melting (breaking the orientational order
between the shells) precedes melting inside the shells. Namely, the
reorientations of shells (jump rotational diffusion) or (with
increasing temperature) free relative rotation of shells take place
before shell structure breaking. This phenomenon is referred to as
orientational melting. Moreover, the study of free relative
rotation of shells may be interesting for nanomechanics \cite{tpb}.

The van der Waals interaction between atoms of neighbor shells in
carbon nanoparticles is considerably weaker than chemical
bonds between atoms inside the shell. So it is natural that
these nanoparticles are possible candidates for orientational melting
\cite{pe}. The possibility of orientational melting of long
two-shell carbon nanotube was discussed \cite{tom0}.
The orientational melting in carbon nanotube bundle was also
theoretically studied \cite{tom1}.

In the present paper the zero temperature energetic characteristics of
two-shell carbon nanoparticle $C_{60}@C_{240}$ are calculated. The
values obtained for barriers of relative rotations of shells are small
enough for free rotation of shells to take place at room temperature. The
orientational melting of this nanoparticle is studied here by
molecular dynamics technique. The definition of orientational melting
temperature is proposed. The corresponding temperature for
nanoparticle $C_{60}@C_{240}$ is calculated

\section{Simulation details}

The following reasons have determined our choice of nanoparticle
shells. The TEM images show that the inner shell of carbon
nanoparticle can have a size that is close to that of
fullerene C$_{60}$ \cite{7,8}. The fullerene C$_{60}$ with $I_h$
symmetry is the smallest fullerene without adjacent pentagons in
its structure. Fullerenes smaller than C$_{60}$ can not be
directly extracted by the use of any solvent from soot, obtained
in arc discharge (see, e. g., Refc. \cite{9,10}). To explain this
fact it was proposed that atoms of fullerenes which belong to two
adjacent pentagons can have chemical bonds with neighbor
fullerenes in soot \cite{11}. For example, chemical bonds between
all neighbor fullerenes are present in solid C$_{36}$ \cite{10a}.
Therefore we consider C$_{60}$ as the smallest inner shell where
the absence of chemical bonds between shells is very probable (it
is a necessary condition for existence of relative rotation of shells). The
single and double bonds lengths of C$_{60}$ used are 1.391 $\AA$ and
1.455 $\AA$, respectively \cite{b22}. We accept the fullerene
C$_{240}$ with $I_h$ symmetry as outer shell of nanoparticle. This
model gives the distance between shells in agreement with
experiment \cite{8} being close to the distance between graphite
planes. Besides, fullerene C$_{240}$ with $I_h$ symmetry have
greater binding energy than fullerenes C$_{240}$ with other
structures \cite{p1}. Several sets of geometric parameters
corresponding to different shapes of fullerene C$_{240}$ obtained
by {\it ab initio} calculations of minima of binding energy
\cite{pe,p2,p3} are used. Different shell shapes $B$, $C$, $D$ and
$E$ were found by optimization of all independent geometric
parameters of fullerene C$_{240}$ with $I_h$ symmetry. The $B$ and $D$ shapes
corresponding to global and local minima found by York
{\it et al} \cite{p2} that are close to sphere and truncated
icosahedron, respectively. Shape $E$ corresponds to the single
minimum found by Osawa \cite{pe}. It is intermediate between
shapes $B$ and $D$. Shape $C$ is rather close to shape $E$.
It corresponds to the minimum found by Scuceria \cite{p3}. The
shape $A$ is obtained by optimization of fewer of
independent geometric parameters so that all atoms of this shape
are arranged on the sphere \cite{p2}.

We describe the interaction between atoms of neighbor shells
by Lennard-Jones potential $U=4\epsilon((\sigma/r)^{12}-(\sigma/r)^{6})$
with parameters $\epsilon=28$ K and $\sigma=3.4$ $\AA$. These
parameters were used for the simulation of solid C$_{60}$
\cite{b20}. The interaction between atoms inside shells are
described by Born potential:

\begin{equation}
U = \frac{ \alpha-\beta}{2} \sum ^{60}_{i,j=1}
( \frac{ ({\bf u}_{i}-{\bf u}_{j}){\bf r}_{ij}}{|{\bf r}_{ij}|} )
^{2}+ \frac{ \beta}{2} \sum^{60}_{i,j=1}({\bf u}_{i}-
{\bf u}_{j})^{2}
\end{equation}

\noindent where ${\bf u}_{i}$, ${\bf u}_{j}$ are displacements of atoms
from equilibrium positions, ${\bf r}_{ij}$ are distances between atoms.
We take $\alpha=1.14 \cdot 10^{3}$ N/m and $\beta=1.24 \cdot 10^{2}$
N/m. Born potential with these values of force constants
gives an adequate internal vibrational spectrum of C$_{60}$
\cite{b21}. Born potential is correct only near the bottom of
potential well. Nevertheless we believe that this potential is
adequate for our simulation because we use it at temperatures
that are one-two order of magnitude less than the temperature of
fullerene destruction.

We studied the orientational melting of nanoparticle
$C_{60}@C_{240}$ with shape $D$ of C$_{240}$ by
molecular dynamics technique. The simulations are performed in
microcanonical ensemble. The equations of motion were integrated
using the leap frog algorithm. We used the integration step
$\tau=6.1\cdot 10^{-16}$ s (about one hundred steps for period of
atoms vibration inside shells). Initially the system has been
brought to the equilibrium during 300-500 ps that is about 30--50
librations of shells. The average fluctuations of the total energy
and temperature of the system fall and flatten out during this
period. Then the system was studied during 100 ps. The average
fluctuations of the total energy of the system were within 0.3 \%
and the average fluctuations of temperature were within 1.3 \%.
The angular velocities of shells change rather slowly to average
the properties of system over the different directions of
angular velocities during one computer experiment. Therefore
all investigated quantities were averaged over 34-46 different
realizations of the systems at the same temperature but with
different random angular velocities of shells corresponding
to their distribution at temperature studied.

\section{Results and discussion}

\subsection{Ground state energetics}

The global and local minima of total nanoparticle energy are found
by optimization of three angles of their {\it relative} orientation.
The total nanoparticle energy includes the energy of interaction
between shells and the energy of shell deformation. We describe the
relative orientations corresponding to minima of total energy in
terms of three angles $\alpha_z$, $\alpha_y$ and $\alpha_x$ of
subsequent rotations of first shell around axes OZ, OY and OX of
coordinate system. The centers of both shells coincide with the
center of coordinate system. The angles $\alpha_z$, $\alpha_y$ and
$\alpha_x$ were measured from the initial orientation shown on Fig.
1. Due to the high $I_h$ symmetry of shells the number of any
equivalent minima (global or local) is 60. Such equivalent minima
correspond to different relative orientations of shells. The
energies of interaction between shells and angles of one of the
orientations corresponding to global and local minima of total
energy of nanoparticle are listed in Table 1.

The energies of interaction between shells calculated here are
slightly less than 16.9 \cite{p4}, 18.57 \cite{p5} and 20.3
\cite{p4} meV/atom obtained using another representations of van
der Waals interaction and are about three times less than
estimation 65.3 meV/atom for graphite \cite{11a}. Note, that the
energy of total interaction between shells is not maximal for
perfect sphere in comparison with other shapes of C$_{240}$
contrary to the assumption of Lu and Yang \cite{p5}.

We observed that the angles of orientations corresponding to global
and local minima are determined by the shape of second shell. For
shapes $C$, $D$ and $E$ of C$_{240}$ the initial relative orientation
of shells (where symmetry axis of shells coincide) corresponds to
global minima of total nanoparticle energy (note, that all these
shapes of C$_{240}$ are close to the truncated icosahedron). Several
global minima for shape $D$ are shown on Fig. 2a. One type of local
minima is found for these shapes of C$_{240}$. For the shape $B$
(which is close to sphere) orientations with coinciding symmetry axes
correspond only to local minima (see Fig 2b). No minima correspond to
such orientations for the "spherical" shape $A$. For the "spherical" shape
of C$_{240}$ two types of local minima are found. The differences
$\Delta E_{loc}$ in total nanoparticle energies between global and
local minima are very small and also determined by the shape of second
shell (see Table 2). The differences $\Delta E_{loc}$ decrease with
decreasing the average deviation $<\Delta R_{i2}>=<|R_{i2}-<R_{i2}>|>$
of second shell from perfect sphere, where $R_{i2}$ is the distance
between an atom of second shell and the center of nanoparticle. Thus
the differences $\Delta E_{loc}$ are small for the "spherical" shape
$A$ and close to sphere shape $B$. The differences $\Delta E_{loc}$
also decrease when the average distance between shells
$h=<R_{i2}>-<R_{i1}>$ approaches the distance $r_{min}$
corresponding to the minimum in pair interatomic potential. This fact
can be explained as follows: the smaller is the difference between $h$
and $r_{min}$ the lesser part of distances $d_{12}$ between two atoms
of neighbor shells corresponds to steeply rising interatomic
potential well. Consequently, the change of distances $d_{12}$ during
relative rotation of shells causes the less change of interaction
energy between shells.

The calculated energies of shell deformation are presented in
Table 2. The influence of shell deformation on the barriers of
relative rotation of shells is studied as an example for barriers
$B_5$ of shell rotation around fivefold axes.
(Barriers $B_5$ were calculated for the relative orientation where
symmetry axes of shells have the same directions). Comparison
of barriers $B_5$ calculated with and without shell
deformation gives a difference less than 1 \% for
all five shapes of C$_{240}$ investigated here. (Note that the
barrier $B_5$ calculated here for the shape $E$ of C$_{240}$ is 12 \%
less than that obtained by Osawa \cite{pe} who used the tandem of molecular
orbital and molecular mechanics calculations). Therefore,
the shell deformations are disregarded here in calculation of
barriers of relative rotation of shells, i.e. lengths of bonds and
angles between bonds inside shells are supposed to be fixed during
intershell rotation. Note that an opposite situation take place, e.g. for
clusters with logarithmic interaction between
particles \cite{4}. In this case, the interparticle interactions
inside shell and between shells are the same and, therefore, the
considering of shells deformation is necessary in calculation of
barriers for rotation. The relative displacement of the centers of
symmetry of shells causes an increase in intershell interaction
energy. Therefore, the common center of symmetry of both shells
is also supposed to be fixed during rotation.

The barriers of relative rotation of shells in the nanoparticles under
consideration are calculated for relative orientations corresponding
to global minima of total nanoparticle energies. It is found that the
values of barriers obtained for rotation are {\it surprisingly small}
(see Table 2). Moreover, these barriers are only several times greater
than barriers $B_a$ in dependencies of interaction energy between {\it
only one atom} of the second shell and the whole first shell vs. angle
of rotation. For example, for the nanoparticle with shape $D$ of
C$_{240}$ the barrier for rotation around fivefold axis is 158.8 K.
Simultaneously, the maximal barrier among the barriers $B_a$ for
different atoms of the second shell is 21.6 K. Detailed analysis
shows that maxima of barriers $B_a$ for individual atoms in the same
shell correspond to {\it different} angles of rotation and so the
dependence of total energy on angle of rotation is {\it essentially
smoothed} (see Fig. 3). Magnitudes of barriers of relative shell
rotation are very sensitive to the shape of C$_{240}$ and decrease
when $<\Delta R_{i2}> \to 0$ and $h \to r_{min}$ (analogously to the
differences $\Delta E_{loc}$ in interaction energies between global
and local minima). Note that the using of spherical shape of
C$_{240}$ leads to significant underestimation of barriers for
rotation.

The radii of shells of nanoparticle $C_{60}@C_{240}$ are very close
to radii of shells of (5,5)@(10,10) two-shell carbon nanotube. It is
interest that barriers for relative rotation of shells per one atom
calculated here for all considered nanoparticles are order of
magnitude less than appropriate barrier
in (5,5)@(10,10) two-shell carbon nanotube calculated by Kwon and
Tomanek \cite{tom0}.

\subsection{Molecular dynamics simulation}

We have investigated by molecular dynamics technique the angular
velocity autocorrelation function of shells, the spectrum of shell
librations, the frequency of shell reorientations, distributions of
Eiler angles of relative orientations of shells, heat capacity of
nanoparticle and barriers in intershell interaction energy
corresponding to shell reorientation events.

The dependence of total energy on temperature is used to calculate
the heat capacity of nanoparticle. In the $30-150$ K temperature
region investigated the heat capacity per one degree of freedom has no
difference from the heat capacity of harmonic oscillator system
within the accuracy of calculation that is less than 5 \%. Only
three degrees of freedom are accounted for relative orientation of
shells. Therefore, as was to be expected, there is not any
peculiarities in the dependency of heat capacity on temperature and
the orientational melting of two-shell carbon nanoparticle has a
crossover behavior.

The dependence of shells reorientation frequency $\nu$ vs.
temperature $T$ is shown on Fig. 4. The jump orientational
intershell diffusion takes place where $kT \ll B_{ef}$, $B_{ef}$ is
an effective energy barrier of reorientation. We interpolate the
reorientation frequency $\nu$ for jump orientational intershell
diffusion at temperatures $30-100$ K by the Arrhenius formula (thick
line on Fig. 4):

\begin{equation}
       \nu = \Omega_0 \exp \left( - \frac {B_{ef}}{kT} \right),
\end{equation}

\noindent where $\Omega_0$ is a frequency multiplier. The fitting
by least square technique gives $B_{ef} = 167 \pm 22$ K
and \mbox {$\Omega_0= 540 \pm 180$ ns$^{-1}$}. Using
a shorter temperature range $T=30-75$ K for interpolation is
found to have only a slight influence on calculated parameters
$B_{ef}$ and $\Omega_0$. Note, that effective barrier of
reorientation $B_{ef}$ is in good agreement (within the accuracy of
calculation) with the minimal $B_{min}$ and average
$B_{av}$ barriers for rotation of shells at zero temperature.
Therefore it is possible to use barriers $B_{min}$ and
$B_{av}$ as effective barrier of
reorientation to estimate the temperature of orientational
melting of carbon nanotubes and nanoparticles with shell
structure.

The exponential increase of
reorientation frequency $\nu$ ends at temperatures $100-150$ Š
and this shows the beginning of free rotation of shells. It
can be shown that the reorientation frequency $\nu$ at temperature
$kT \gg B_{ef}$ can be estimated by the expression

\begin{equation}
\label{3}    \nu = \frac {n}{2 \pi} \sqrt \frac {3kT (I_1 + I_2)}{I_1I_2}
\end{equation}

\noindent where $n$ is an average number of reorientations over
the period of relative shell rotation ($n \approx 5$), $I_1$ and $I_2$ are
moments of inertia of 1-st and 2-nd shells, respectively.
The dependence of reorientation frequency on temperature defined by
Eq. (\ref{3}) is shown on Fig. 4 by thin line.

The prominent smooth distributions of Eiler angles of relative
orientations of shells (Fig. 5), the disappearance of maxima in
the angular velocity autocorrelation function of shells (Fig. 6)
and in the spectrum of shell librations (Fig. 7) confirm that the
free rotation of shells determines the thermodynamical behavior
of the nanoparticle at temperatures greater than \mbox{140 Š.}

The temperature dependence of the "experimental" barriers $B_{re}$
in intershell interaction energy corresponding to shell
reorientations events is shown on Fig. 8. The barriers $B_{re}$ are
averaged over all observed reorientation events at corresponding
temperature (30--70 reorientation events for each temperature
investigated from range $T=40-55$ K and 200--600 reorientation
events for each temperature investigated from range $T=70-150$ K). At
temperatures 30--100 K, where the jump orientational intershell
diffusion takes place, barrier $B_{re}$ is in agreement (within the
accuracy of calculation) with the minimal barrier $B_{min}$ for
rotation of shells at zero temperature and with effective barrier
of reorientation $B_{ef}$. At temperatures $100-150$ Š, where
free rotation of shells begins, the magnitude of "experimental"
barrier $B_{re}$ grows and the increase of this barrier $\delta
B_{re}$ runs to 50 K at temperature 154 K. The increase $\delta
B_{re}$ is greater than the dispersion $\Delta B_{av}=10$ K of
barriers for rotation of shells at zero temperature. Consequently,
the increase $\delta B_{re}$ of the barrier can not be explained
by climbing over the barrier with increasing temperature not only
at their lowest point. Therefore, we believe that the increase
$\Delta B_{re}$ of the barrier is the result of shell deformation.
The increase of the dispersion $\Delta B_{re}$ of "experimental"
barrier with increasing temperature (see Fig. 9) also indicates
the influence of shell deformation on this barrier. Note that
energy of shell deformation is three order of magnitude greater
than the increase $\delta B_{re}$ of "experimental" barrier in the
result shell deformation. (The energies calculated of shell
deformation $E_d$ are in agreement with virial theorem
$E_d=kT(3N-6)/2$, where $N$ is number of atoms in a shell.)

Phenomena of order breaking in the systems with finite number of
particles occur as a rule at some temperature range. This leads to
problems in attempting to define the temperature of order breaking
(see, e.g., Ref. \cite{att} and references herein). In the majority of
cases a melting of clusters occurs with change of cluster structure.
That is the system become spend time with increasing temperature not
only in the ground state but also in states with the structure
corresponding to other minima in potential energy of system.
A mesoscopic system can fluctuate between different states
separated by an energy barrier: the solid state corresponding to the
ground state of the system and the liquid-like state corresponding
to other minima in potential energy
(see, e.g., Refc. \cite{belo}). To characterize such a system
at its melting the quantity $K(T)=\gamma_l/\gamma_s$ was introduced
\cite{ber}, where $\gamma_l$ and $\gamma_s$ are the probabilities
that the system is at the temperature $T$ in liquid-like state and
solid state, respectively. In this case the temperature $T_c$
corresponding to $K(T_c)=1$ may be considered as melting
temperature.

The situation is different for the considered nanoparticle
$C_{60}@C_{240}$. The reorientations of shells are transitions between
states corresponding to {\it equivalent} minima of potential energy. That
is the structure of system does not change during orientational
melting. To characterizes the melting of such systems, where
diffusion occurs during the melting but the structure of the system
does not change, we introduce the quantity $K(T)=\nu_t/\omega_t$,
where $\nu_t$ is the frequancy of transitions between equivalent
minima of potential energy and $\omega_t$ is the frequancy of such
oscillation wherein movement of particles directed along the path of
transition. For considered systems we propose to define the
temperature $T_c$ corresponding to $K(T_c)=1$ as the melting
temperature. In this case a half of appropriate oscillations
follows by transition to equivalent minima. Note, that proposed
definition corresponds to a temperature of short order breaking
and has no anology with phase transitions in macroscopic systems
(contrary to above definition for the systems with transitions
between inequivalent minima).

The appropriate quantity characterizing the orientational melting
of nanoparticle $C_{60}@C_{240}$ is $K_o(T)=\nu/\omega$,
where $\nu$ is the shell reorientation frequency and $\omega$ is
the frequency of relative librations of shells.
The equality $K_o=1$ implies that a half of relative
librations of shells begins in one minimum in dependence of
potential energy of nanoparticle on angles of relative shell
orientation and ends in neighbor equvalent minima. The estimation
with the help of simulation performed gives the temperature $T_c$ of
orientational melting for nanoparticle $C_{60}@C_{240}$ with
shape $D$ of the second shell $T_c \approx 60$ K ($T_c$ corresponds
to $K_o(T_c)=1$). Here the frequency of relative
librations of shells is determined from the maximum in the spectrum
of shell librations (see Fig. 7).

As we have shown, the barriers for rotation are very sensitive to the
shape of shells. Therefore, the realization of possible orientational
melting in many-shell nanoparticles is determined by their shape.
The nanoparticles obtained in arc discharge are faceted in shape
\cite{6,12}. However, their shape changes to almost spherical
one when they are subjected to very strong electron irradiation in a
high-resolution electron microscope \cite{8,11a,13}. Accurate
{\it ab initio} calculation of geometric parameters of large shells
is necessary for performance of theoretical studies of possible
orientational melting of many-shell nanoparticles. Nevertheless, the
theory does not provide accurate coordinates. Some works predict
that many-shell nanoparticles are faceted \cite{p1,p4}, some that
they are spherical \cite{p5,p6}, and some the transition from
faceted to spherical shape for shells containing more than 3500
atoms \cite{per}. The calculations have also shown that the faceted
nanoparticles transform to spherical under high temperature
\cite{p4,p7}. Therefore, the barriers for rotation may {\it decrease}
with {\it increasing} of temperature due to change of shell
structure. Thus it is found that temperature of orientational melting
$T_c \approx 60$ K for two-shell carbon nanoparticle is at
least one order of magnitude less than the temperature of total melting.
Analogously orientational melting can occur also in many-shell
nanoparticles and short many shell nanotubes \cite{tbp}.

The carbon nanoparticles with shell structure are not the single
example of different types of atom interaction inside shell and
between shells. A two-shell spherical nanoparticle from $MoS_2$
was produced \cite{p6}. We believe that orientational melting can
also take place in nanoparticles from this and analogous sandwich
materials ($MX_2$, $M=Mo,W$, $X=S,Se$).

The orientational melting in a single nanoparticle may be revealed
by IR or Raman study of the temperature dependence of width of
spectral lines. The last must have Arrhenius-like contribution in
reorientational phase (analogously to the behavior in plastic
crystals, see, e.g., Ref. \cite{15} and references
herein). Moreover, this study can give the estimation of
reorientational barriers. Besides, NMR line narrowing can be observed
in reorientational phase.

\section*{Acknowledgements}

This work was supported by grants from Russian Foundation of
Basic Researches, Programs
"Fullerenes and Atomic Clusters" and "Surface and Atomic Structures".\\

\newpage
\vspace{1cm}

Table 1.

The energies $E_{int}$ of interaction
between nanoparticle shells and one of the relative orientations of
shells corresponding to the global and local minima of total
nanoparticle energy; $\alpha_z$, $\alpha_y$ and $\alpha_x$ are the
angles of subsequent rotations of inner shell from initial orientation
around axes OZ, OY and OX, respectively.\\

\begin{tabular}{ccccc}
\hline
Shape & $E_{int}$, & $\alpha_z$ & $\alpha_y$ & $\alpha_x$ \\
 & (meV/atom)  & (in radians) & (in radians) & (in radians) \\
\hline
$A$ & 15.034 & 0.0819 & 0.1452 & 0.0540 \\
$A$ & 15.033 & 0.2495 & 0.8128 & -0.0081 \\
$A$ & 15.032 & 0.6283 & 0.4634 & 0.0 \\
\hline
$B$ & 15.124 & 0.6283 & 0.4634 & 0.0 \\
$B$ & 15.101 & 0.0 & 0.0 & 0.0 \\
\hline
$C$ & 15.180 & 0.0 & 0.0 & 0.0 \\
$C$ & 15.098 & 0.6283 & 0.4634 & 0.0 \\
\hline
$D$ & 13.819 & 0.0 & 0.0 & 0.0 \\
$D$ & 13.777 & 0.6283 & 0.4634 & 0.0 \\
\hline
$E$ & 15.166 & 0.0 & 0.0 & 0.0 \\
$E$ & 15.061 & 0.6283 & 0.4634 & 0.0 \\
\hline
\end{tabular}

\newpage
\vspace{1cm}

Table 2.

The characteristics of second shell shape: the
average deviation of second shell from perfect sphere $<\Delta
R_{i2}>$ and the difference between average intershell distance
$h$ and the distance $r_{min}$ corresponding to the minimum in pair
interparticle potential $l=h-r_{min}$; the differences
$\Delta E_{loc}$ in total energies of nanoparticle between global and local
minima; the minimal and average barriers for
rotation $B_{min}$, $B_{av} \pm \Delta B_{av}$, respectively,
where the barrier $B_{av}$ is averaged over all directions of
rotation axis and $\Delta B_{av}$ is its dispersion; the average
energies of shell deformation $E_{d1} \pm \Delta E_{d1}$ and
$E_{d2} \pm \Delta E_{d2}$ for first and second shells,
respectively, where the energies $E_{d1}$ and $E_{d2}$ are
averaged over all relative orientations of shells and $\Delta E_{d1}$
and $\Delta E_{d2}$ are their dispersions.\\

\noindent
\begin{tabular}{cccccccc}
\hline
Shape & $<\Delta R_{i2}>$ & $l$ & $\Delta E_{loc}$ &
$B_{min}$ & $B_{av} \pm \Delta B_{av}$ & $E_{d1} \pm \Delta E_{d1}$
& $E_{d2} \pm \Delta E_{d2}$\\
 & ($\AA$) & ($\AA$) & ($^o$K) & ($^o$K) & ($^o$K) & ($^o$K) & ($^o$K)\\
\hline
$A$ & -0.245 & 0.0 & 3.2; 5.5 & 19.0 & 20.5 $\pm$ 0.8&
2.09 $\pm$ 0.02 & 34.56 $\pm$ 0.12\\
\hline
$B$ & -0.258 & 0.057 & 76.7 &  82.9 & 122.1 $\pm$ 12.1 &
1.62 $\pm$ 0.07 & 29.98 $\pm$ 0.50\\
\hline
$C$ & -0.289 & 0.152 & 287.4 & 349.3 & 363.1 $\pm$ 8.8&
2.17 $\pm$ 0.26 & 18.19 $\pm$ 0.42\\
\hline
$D$ & -0.119 & 0.244 & 144.4 &  160.3 & 177.3 $\pm$ 9.6 &
3.75 $\pm$ 0.20 & 34.40 $\pm$ 0.55\\
\hline
$E$ & -0.299 & 0.147 & 368.3 & 441.2 & 459.9 $\pm$ 12.9 &
4.58 $\pm$ 0.44 & 13.78 $\pm$ 0.38\\
\hline
\end{tabular}

\newpage
\begin{center}
{\bf Captions for illustrations.}
\end{center}

{\bf Fig. 1.} The fragments of two shells (shape $D$ of second
shell) at their initial orientations. OX, OY and OZ are axes of
coordinate system. One fivefold axis of each shell is aligned with
the axis OZ. One of the closest to axis OZ atoms of first and second shells
(shown by black circles) lie in plane OXZ. This
fixes the orientation of axes OX and OY.

{\bf Fig. 2.} The dependencies of binding energies for interaction
between shells of nanoparticle on their relative orientation.
$\alpha_z$ and $\alpha_y$ are the angles of subsequent rotations of inner
shell from initial orientation around axes Z and Y
respectively. The angle of rotation around axis X is fixed
equal to zero. a) shape D of second shell;
b) shape B of second shell;

{\bf Fig. 3.} Interaction energies between
first shell of nanoparticle and groups of atoms of second
shell with shape $D$ {\it vs.} angle $\alpha_z$ of rotation of inner
shell from initial orientation around axis Z. An each group
include all atoms with the same dependencies of interaction
energy $E_a$ between this atom and the first shell on angle of
rotation. The curves corresponding to all 25 groups of atoms
with different dependencies $E_a$ for individual atom are shown
by thin lines (23 groups from 10 atoms and 2 groups from 5
atoms). The dependence of total interaction energy between shells
on angle $\alpha_z$ is shown by bold line. All energies are measured
from their minima.

{\bf Fig. 4} The dependence of shells reorientation frequency $\nu$
on temperature $T$ in Kelvin degrees. The interpolation
by the Arrhenius formula at $kT < B_{re}$ is shown by thick line.
The estimation at $kT > B_{re}$ is shown by thin line.

{\bf Fig. 5} The distributions of Eiler angles $\theta$, $\psi$ and $\phi$
of relative orientations of shells at temperatures 21 K, 36 K and 140 K
are shown by dotted lines, thin lines and thick lines respectively;
a) the distribution of angle $\phi$; b) the distribution of angle $\theta$;
c) the distribution of angle $\psi$.

{\bf Fig. 6} The angular velocity of autocorrelation
function of the first shell at temperatures 21 K, 36 K and 140 K
are shown by dotted lines, thin lines and thick lines respectively.

{\bf Fig. 7} The spectrum of shell librations at
temperatures 21 K, 36 K and 140 K
are shown by dotted lines, thin lines and thick lines respectively.

{\bf Fig. 8} The dependence of the "experimental" barriers $B_{re}$
in intershell interaction energy on temperature $T$.

{\bf Fig. 9} The dependence of the dispersion $\Delta B_{re}$ of barriers
in intershell interaction energy on temperature $T$.
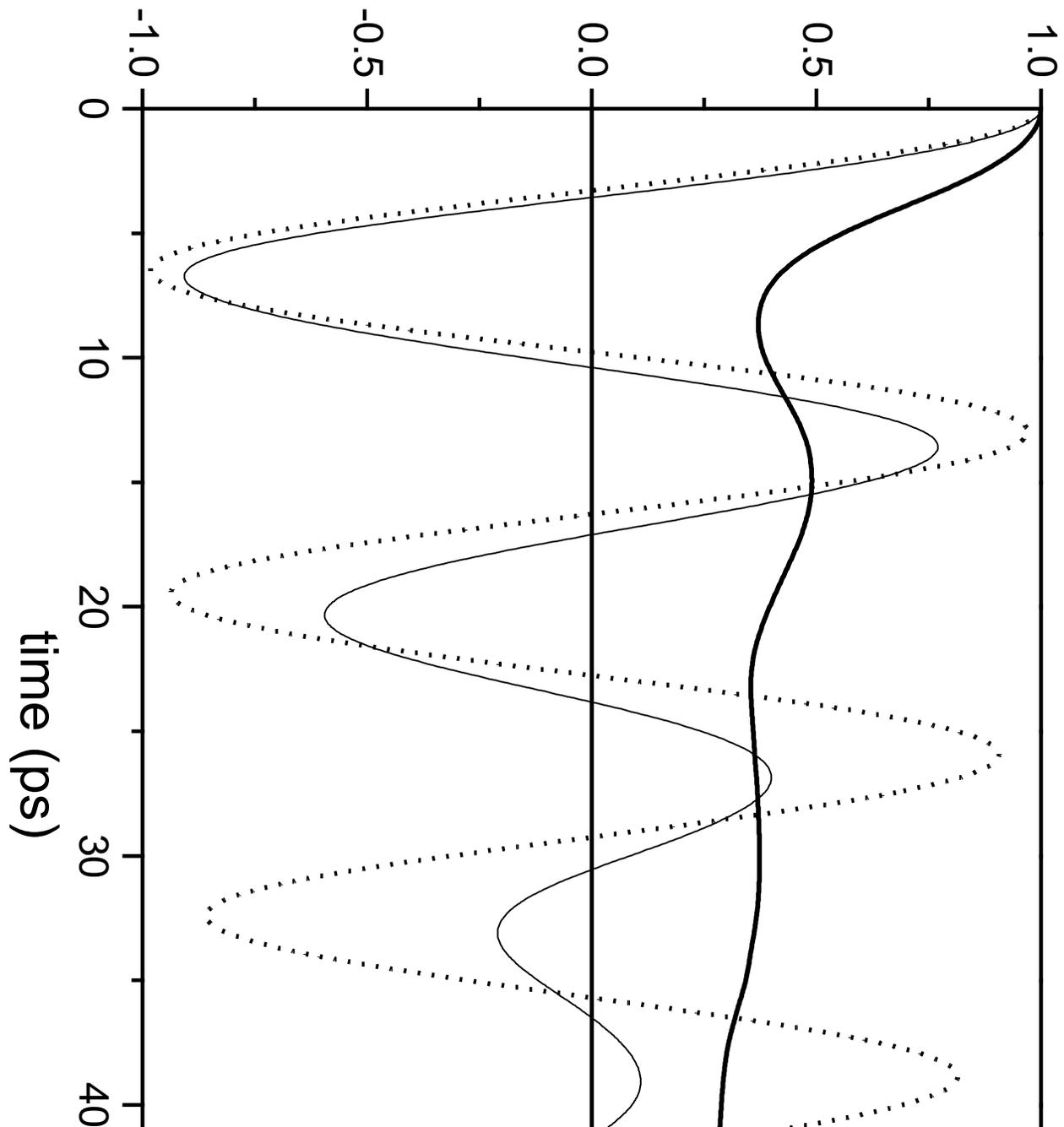
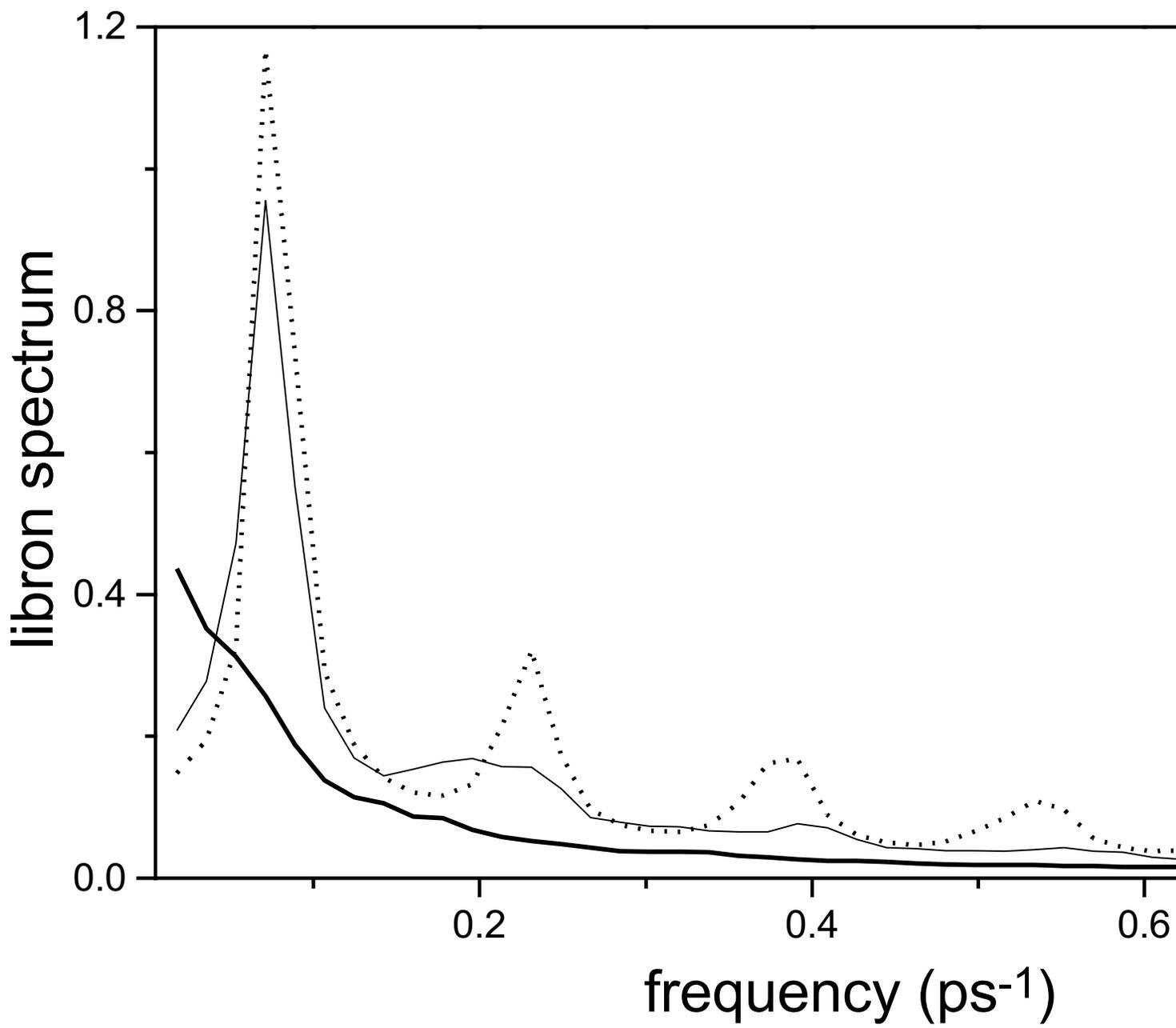
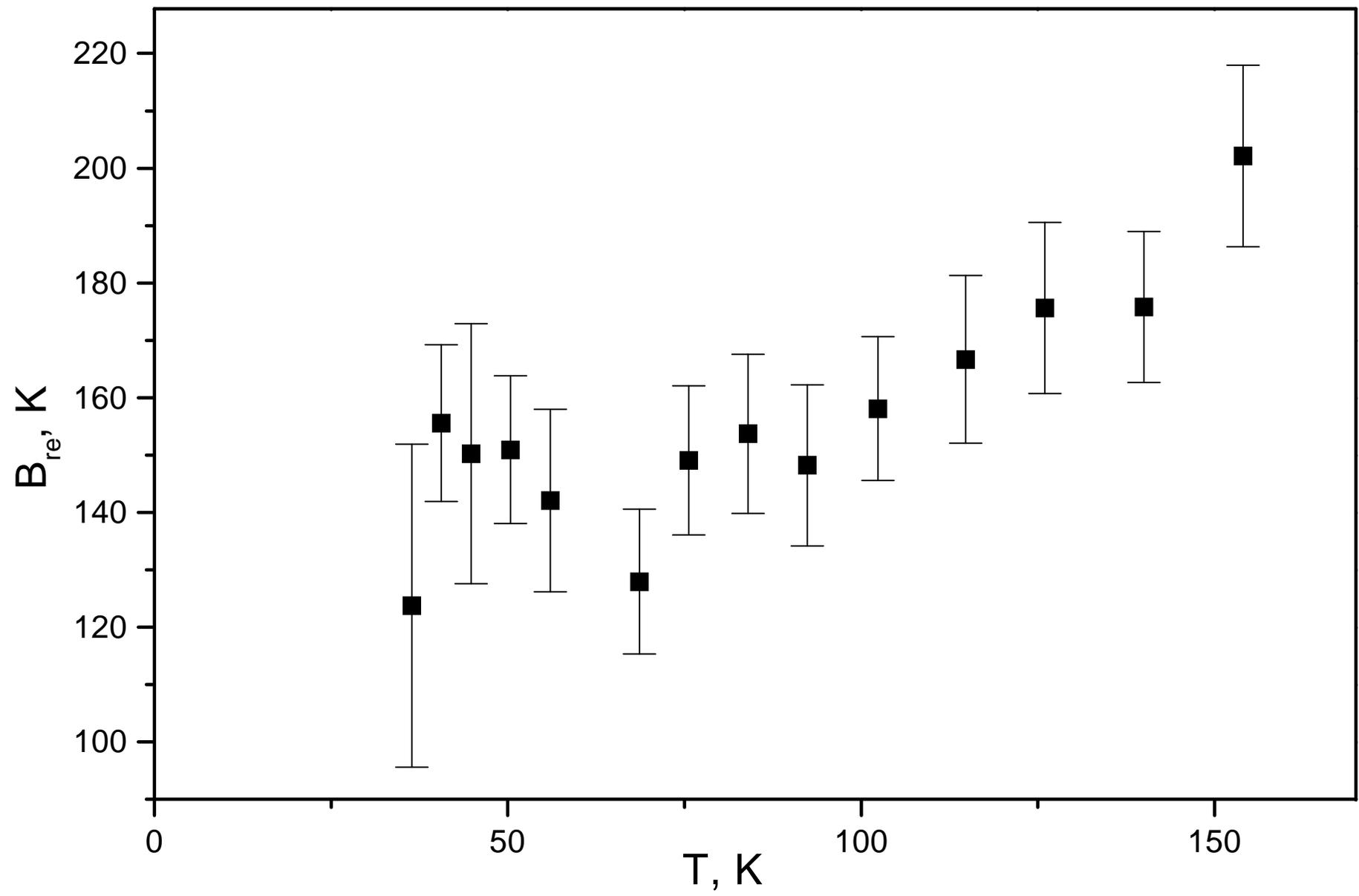
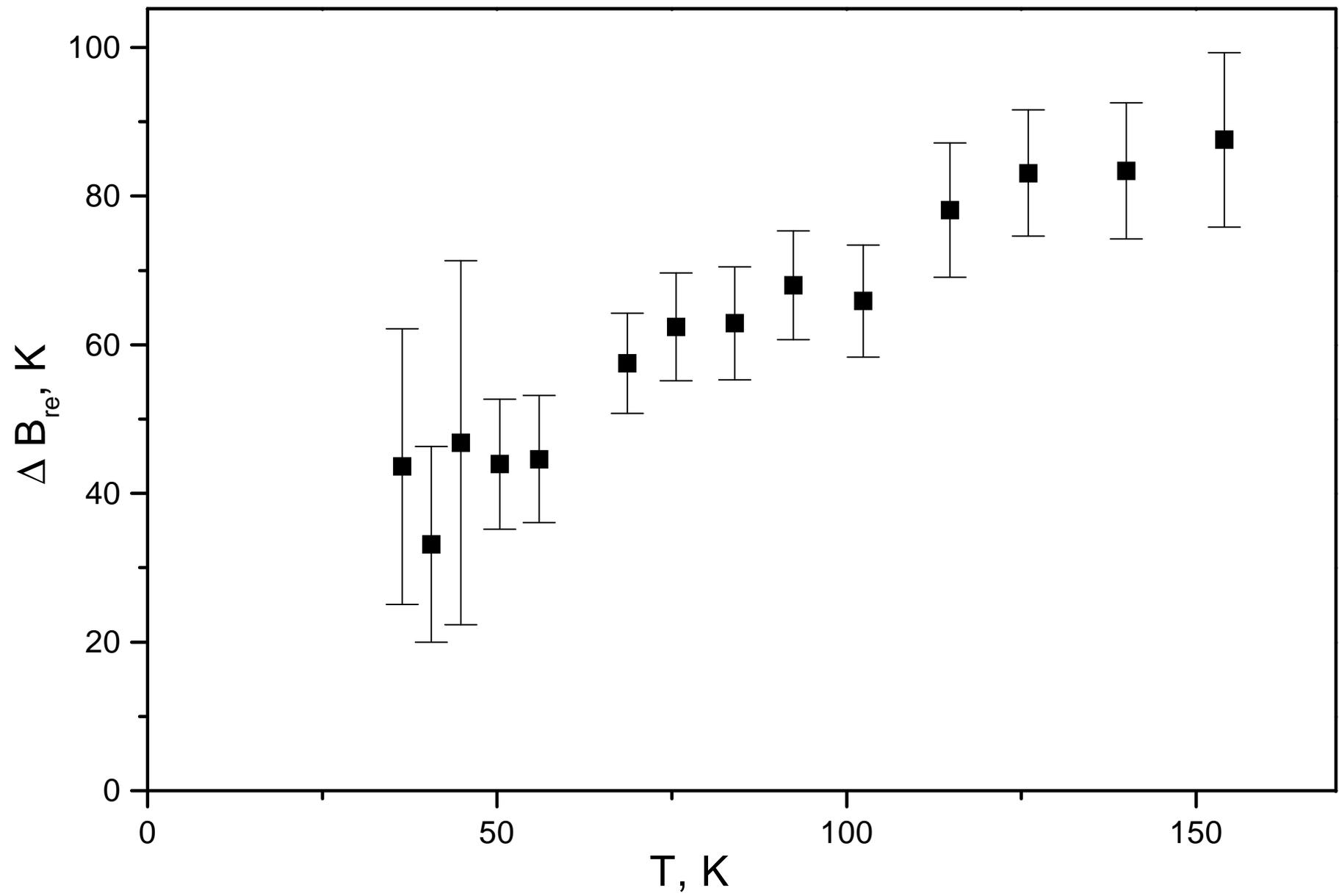
\end{document}